\documentclass[a4paper,12pt]{revtex4}

\newcommand{\ket}[1]{\ensuremath{|#1 \rangle}}
\newcommand{\bra}[1]{\ensuremath{\langle #1|}}

\newcommand{\av}[1]{\ensuremath{\langle #1 \rangle}}

\newcommand{\iden}{\ensuremath{\openone}}

\newcommand{\ot}[2]{\ensuremath{\left( \begin{array}{c} #1 \\ #2
\end{array} \right)}}

\usepackage{amssymb}
\usepackage{graphicx}
\begin{document}
\title{Hybrid hypercomputing: towards a unification of quantum and classical computation.}
\author{Clare Horsman and William J. Munro}
\address{Hewlett-Packard Laboratories, Filton Road, Bristol BS34 8HZ}

\begin{abstract}
We investigate the computational power and unified resource use of hybrid quantum-classical computations, such as teleportation and measurement-based computing. We introduce a physically causal and local graphical calculus for quantum information theory, which enables high-level intuitive reasoning about quantum information processes. The graphical calculus defines a local information flow in a computation which satisfies conditions for physical causality. We show how quantum and classical processing units can now be formally integrated, and give an analysis of the joint resources used in a typical measurement-based computation. Finally, we discuss how this picture may be used to give a high-level unified model for hybrid quantum computing.
\end{abstract}

\maketitle
\section{Introduction}


The standard approach to quantum computing treats quantum and classical information processing and communication as essentially two separate systems. Quantum and classical computations are described using different formalisms, and consume separate resources. While they are frequently combined, for example in protocols such as teleportation \cite{telepo} and dense coding \cite{denco} the interface between them is at best not formalised and at worst entirely obscured by standard presentations. Some `translation rules' exist between quantum and classical \cite[ch1]{nandc}, and we can formally define the complexity of specific quantum algorithms compared to their classical counterparts \cite{quantcomplex}, but we still lack an underlying model of computation and resources for both that would enable direct comparisons to be made. It is even the case (although often overlooked) that the lack of a proof for the extended Church-Turing thesis means that it is  still an open question whether quantum computers are in fact significantly more powerful than classical computers \cite{Bennett94strengthsand}. Given the potentially important applications of quantum computing methods in areas such as cryptography and quantum simulation \cite{crypto,sim}, it is actually a significant question whether we should deploy our resources to implement them by physically building quantum computers, or by investigating new classical algorithms to run on current hardware. Without a unified framework for quantum and classical computations, it is hard to see how we could begin to answer questions such as this with any degree of certainty.

The problem becomes particularly acute when we consider hybrid computing schemes such as measurement-based quantum computing, where the feed-forward of classical measurement results is an integral part of the quantum design \cite{danmbqc,owqc}. Consideration of the resources used during such a computation tends to concentrate solely on the quantum resources required (generally in the form of a cluster state \cite{cluster}), with classical measurement and communication being taken as essentially resource-free. This leaves open problems both of the resources required to physically implement these computations, and also fundamental questions on the computability power of such algorithms. Further problems arise if we consider hybrid cryptographic schemes where formally provable security is desired. The lack of formal links between the quantum and classical elements gives a potential area of uncertainty in the security proofs for such systems.

In this paper we present a candidate for a framework in which to analyse hybrid computation by fully integrating the quantum and classical resources and processes used. This takes the form of a graphical calculus based on the logical Heisenberg picture \cite{gottesman,dh,mevdh}, closely related to the stabilizer formalism that is typically used to describe measurement-based quantum computation \cite{dan}. The framework integrates quantum and classical elements directly, and gives an intuitively simple set of formally exact reasoning tools for hybrid information theory. There is a local flow of information, enabling data to be tracked causally through a computation, including through what are typically considered  to be simply `classical' communication channels. Furthermore, this flow satisfies all reasonable assumptions of causality in physical systems, such as one-way action in time.  We will find that this framework enables us to describe the resource use of a measurement-based computation in terms of a single set of resources, and gives us a much clearer idea both of the physical processes taking place during thes computation and the logical dependency structure of various algorithms. We will finish by discussing how the framework may be abstracted to give us a formal computability model for hybrid computation.

\section{Quantum information processing}

The basic component of a quantum information processing system is a two-level quantum state called a quantum bit or qubit. As with classical two-level systems (bits) the levels are usually denoted as the computational `0' and `1', and a measurement of a qubit in the computational basis will return one of these values with a probability depending on the quantum state. Unlike classical bits, however, outside a measurement context a qubit can be in a `mixture' of the 0 and 1 states (which are written $\ket{0}$ and $\ket{1}$), called a \emph{superposition}. Qubits can also maintain correlations that are stronger than those accessible to classical systems, known as \emph{entanglement}. Entanglement can be thought of as `spatially extended' superposition: the joint state of, for example, a system of two qubits is in a superposition, which then gives rise to entanglement between the qubits individually.

Quantum computations are significantly more sensitive to the resources required to complete them than their classical counterparts. Classical information can be freely read, duplicated and deleted; for example, if two copies of a bit are required for an algorithm, a single one can be created and then cloned. By contrast, quantum information is subject to strict theorems on \emph{no cloning} and \emph{no deleting} \cite{clone1,nodel}. No cloning states that we cannot start with a qubit in a given quantum state, interact it with another, and then come out with two copies of the original state (this does not, of course, preclude two qubits being \emph{prepared} in the same, known, state). No deleting is the reverse process: we cannot take a given qubit and change its state using only unitary evolution (that is, quantum evolution without classical measurement) to a `blank' state without transferring the state information elsewhere. These results are for general quantum states; there are specific states which can be cloned and deleted, which however turn out to be indistinguishable from classical `bit' states. Both no cloning and no deleting are closely related to the readability problem for quantum information: in general, measuring the state of a qubit changes it (it will, in general, change from a superposition to a definite value). This again contrasts with classical bits which can be easily read without destruction of their state. 
 
Using these building blocks, information processing in quantum systems can be categorised broadly into reversible and irreversible computations, dependent crucially on the ordering of quantum and classical elements in the process. Reversible computations use unitary gates acting on coherent quantum systems throughout the computation, with only the final readout of the process result yielding classical data. This has traditionally been the framework used for quantum computing, and includes the circuit and quantum walks models \cite{deutschct,skw}. By contrast, the more recently-devised method of measurement-based quantum computing (MBQC) includes irreversible measurements and classical data during the computation itself as well as for the readout. A coherent and highly-entangled quantum state called a \emph{cluster state} is prepared at the beginning of the computation, and then the algorithm is implemented through a time-ordered pattern of measurements on individual qubits or qubit blocks. The exact type of future measurements in the algorithm will depend  on the outcome of previous measurements on the cluster. At the end of the process the entanglement in the cluster state has been used up (it is therefore considered to be the resource of the computation), and the result is read out from previously-designated `output' elements of the original cluster. 

One fundamental measurement-based protocol that we will use frequently (although historically pre-dating the development of MBQC) is teleportation. In teleportation, a qubit in an unknown state is transmitted using two bits of classical information and previously shared entanglement. Essentially the quantum state is converted to classical data, transmitted, and then converted back to quantum, within the framework of two shared entangled qubits. Various teleportation-type schemes are often found in measurement-based computations, as a way of transferring a quantum state (including its entanglement properties) from one part of the cluster to another.

\section{Quantum mechanics in pictures}



There has recently been a significant amount of research into graphical calculi for information flow in quantum processes. This has largely centred on the category-theoretic work of Coecke and Abramsky \cite{cat,kqm}. The calculi abstract out the structure of quantum information processes, and present a high-level framework for reasoning about information flow in systems, generally containing both bits and qubits. There is now a considerable menagerie of diagrammatic representations for the mathematical structure of quantum mechanics (eg. \cite{gc1,gc2}), and it is with some trepidation that we intend to advance yet another. The difference lies in the perspective from which the present calculus has been developed. Previous diagrammatic representations have concentrated on the mathematical, logical and information-theoretic structures of information flow. The present work is concerned rather with the \emph{physical} and \emph{causal} flows of information during quantum computation. These two broad areas are by no means mutually exclusive, but are sufficiently dissimilar to cause difficulties if one attempts to answer questions of interest to one using the representations of the other. A particular example of this comes in the Coecke and Abramsky calculus where there are logical information flows that travel backwards in time. If we want to ask about how information \emph{physically} gets between points in a computation, then such a flow picture will present serious problems. This present calculus has been developed specifically from such a ``physics'' rather than ``computer science'' perspective, with the hope of adding to our integrated understanding of the logical and physical nature of quantum information processing.

\subsection{The logical Heisenberg formalism}

We will first describe the formalism of quantum theory on which the graphical calculus will be based. This is a Heisenberg-picture formulation of quantum mechanics, where operators rather than states give the evolution of a system. The formalism is mathematically entirely equivalent to more standard pictures of quantum mechanics, the only difference being that there is no non-unitary evolution within the system. It was developed by Deutsch and Hayden following Gottesman \cite{gottesman,dh,mevdh}, and can be viewed as a generalised form of Gottesman's stabilizer theory. The basics of this formalism are the Hilbert-Schmidt space \emph{descriptors} for each qubit, from which we can generate the set of all operators on that qubit. These are the Heisenberg operators which track the evolution of the qubit systems through the computation. The initial state for descriptors is defined as the $\ket{0}$ (unentangled) state, which for qubit $a$ in an $n$-qubit system is written 
$$ \mathbf{q}_a = \ot{q_{ax}}{q_{az}} = \iden^{\otimes n-a} \otimes \ot{\sigma_x}{\sigma_z} \otimes \iden^{\otimes a-1}$$

\noindent (In its fullest form we should also have a $y$-component to $\mathbf{q}_a$, but it is not independent of the other components: $q_{ay} = q_{ax}q_{az}$. We therefore do not need to track it separately).

The descriptor formalism is unique in quantum mechanics in that the individual qubit descriptors contain locally all the information available to any joint system
$$ q_{abij} = q_{ai} \ q_{bj} $$

The descriptors evolve only when the specific qubit is subjected to a (single- or multiple-qubit) gate, so remote operations do not change the qubit. Some useful single-qubit gates are the NOT gate,
$$ \mathbf{q}(t_1) = \ot{q_x(t)}{- q_z(t)}$$

\noindent and the Hadamard, which in the Schr\"{o}dinger picture acts as
\begin{eqnarray*} \ket{0} & \rightarrow & \frac{1}{\sqrt{2}}(\ket{0}+\ket{1}) \\
\ket{1} & \rightarrow & \frac{1}{\sqrt{2}}(\ket{0}-\ket{1}) \end{eqnarray*}

\noindent and in descriptor terms is written
$$ \mathbf{q}(t_1) = \ot{q_z(t)}{q_x(t)}$$

%

Another example of a frequently used gate is the controlled-NOT (CNOT) operation between two qubits. A CNOT operation between a control and target qubit performs a NOT on the target if the control is 1, and does nothing if the control is 0. In the descriptor formalism this is written
\begin{eqnarray} \mathbf{q}_1(t_1) & = & \ot{q_{1x}(t) \ q_{2x}(t)}{q_{1z}(t)} \nonumber \\
\mathbf{q}_2(t_1) & = & \ot{q_{2x}(t)}{q_{1z}(t) \ q_{2z}(t)} \label{CNOT}\end{eqnarray}

Entanglement in this formalism is given by a non-trivial dependency of a descriptor outside the space that it occupied at time $t_0$ \cite{mevent}.

The relationship with the standard, Schr\"{o}dinger, formalism of quantum mechanics is given most easily through the density matrix, which can be written for two qubits as
\begin{equation}\rho_{12} = \sum_{i,j=0}^3  \av{q_{1i}q_{2j}} \ \sigma_i \otimes \sigma_j\label{expected} \end{equation}

\noindent where $q_{ay} = q_{ax}q_{az}$.

We can use this relation with the density matrix to simplify the formalism from this rather unwieldy state. We can see from (\ref{expected}) that in terms of measurement outcomes, it is the expected values of the descriptors which are important. These expectation values in (\ref{expected}) are defined with respect to the fixed Heisenberg state $\ket{0}$, which for the Pauli operators gives
$$ \av{\sigma_x} = \bra{0} \sigma_x \ket{0} = \av{\sigma_y} = 0 \ \ , \ \ \ \av{\sigma_z} = \av{\openone} = 1$$

\noindent Recalling the property of the Pauli operators that $\sigma_i \sigma_j = \epsilon_{ijk} \sigma_k$ for $i,j,k \in \{0,1,2,3\}$ (where $\epsilon_{ijk}$ is the Levi-Civita symbol), we therefore have the relations
$$ \av{\sum_{ijkl} (a_i \sigma_x)(b_j \sigma_y)(c_k \sigma_z)(d_l \openone)} = \av{\sum_{ijkl} (a_i \sigma_x)(b_j \sigma_x)(c_k \openone)(d_l \openone)}$$

\noindent This means that we do not need separate $\sigma_y$ or $\sigma_z$ operators in the description -- wherever they occur, they can be replaced by $\sigma_x$ and $\openone$ respectively, without changing any of the current or future predictions of the formalism.

These are now all the elements that are needed to transfer to a graphical calculus. There is an isomorphism between the formalism described and the calculus we will give in the next section, so that anything proven using the calculus can also be proven in the Deutsch-Hayden picture. As this is also isomorphic to the standard quantum formalism, the calculus will form a complete proof net for quantum propositions.

\subsection{Yet another graphical calculus}

From the previous section we can pick out the elements that we need to translate graphically. The representation has as its basic elements qubits, which have two `slots' representing the $x$- and $z$- components of the descriptor. In these slots we need to put representations of our two basic operators, $\sigma_x$ and $\iden$. The representations must satisfy the properties $\sigma_x . \iden = \sigma_x$ and $\sigma_x^2 = \iden$. As this is a Heisenberg formalism, we start from out initial `zero state' (we can take this by convention to be $\ket{0}$) and evolve from there using gates. A `0' qubit is given by
\vspace{0.2cm}
\begin{center}
 \includegraphics[height=0.8cm]{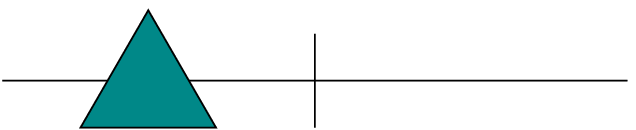}
\end{center}

\vspace{0.2cm}

\noindent The left side represents $q_x$ and the right $q_z$. $\sigma_x$ is given by a triangle, and $\iden$ by a space. We can now perform a NOT gate to gain the `1' state of the computational qubit basis. In the descriptor formalism this acts as $q_x \rightarrow q_x$, $q_z \rightarrow -q_z$, so we represent it (time flows from left to right):

\vspace{0.2cm}
\begin{center}
 \includegraphics[height=0.8cm]{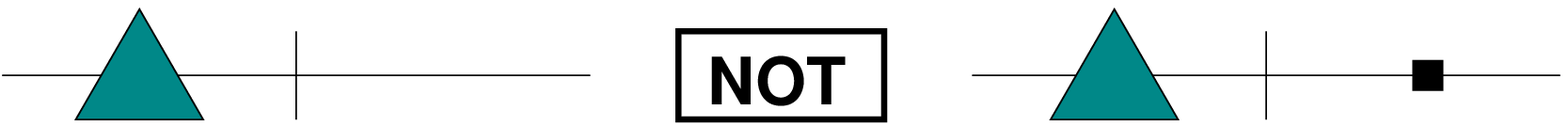}
\end{center}
\vspace{0.2cm}

%

\noindent The final single-qubit operation we are interested in is a Hadamard transform, which swaps $q_x$ and $q_z$ compoents, hence left and right on a qubit:

\vspace{0.2cm}
\begin{center}
 \includegraphics[height=0.8cm]{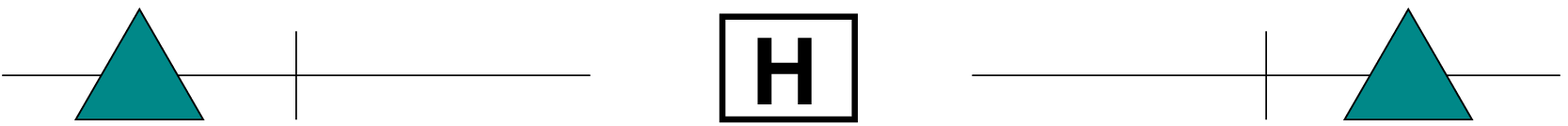}
\end{center}
\vspace{0.2cm}

\noindent A qubit in an unknown state is represented as

\vspace{0.2cm}
\begin{center}
 \includegraphics[height=0.8cm]{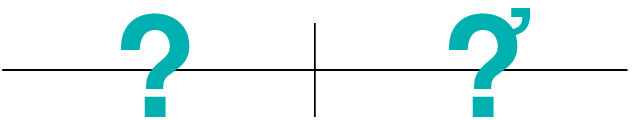}
\end{center}
\vspace{0.2cm}

\noindent Finally we have composition rules for combining known elements within a qubit. We will represent operators on different subspaces of the full Hilbert space by different colours, so that only triangles of the same colour compose, and different colours commute:

\vspace{0.2cm}
\begin{center}
 \includegraphics[height=6cm]{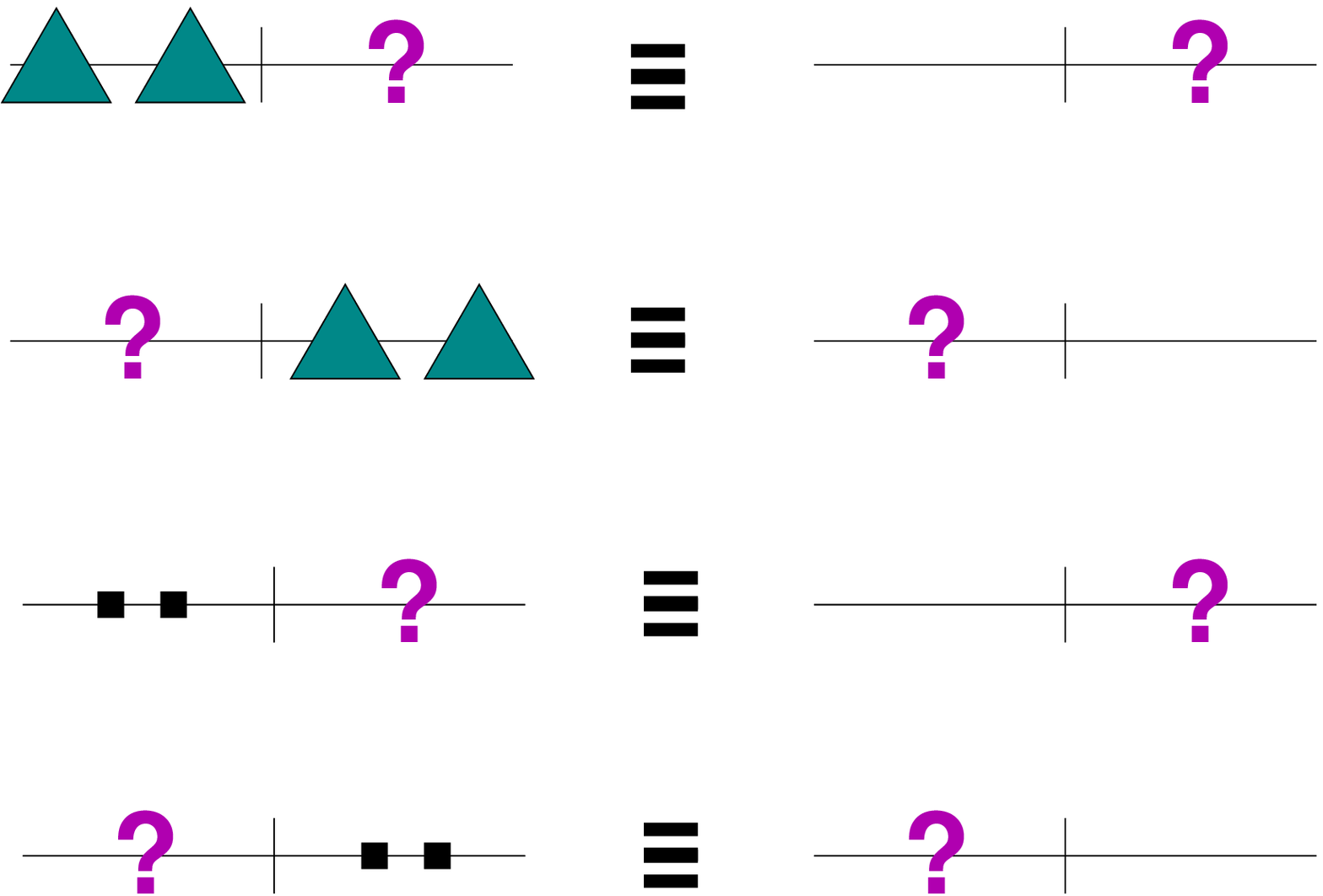}
\end{center}
\vspace{0.2cm}

\noindent Using two qubits in unknown states we can define the general actions of the two-qubit CNOT gate:

\vspace{0.2cm}
\begin{center}
 \includegraphics[height=2.1cm]{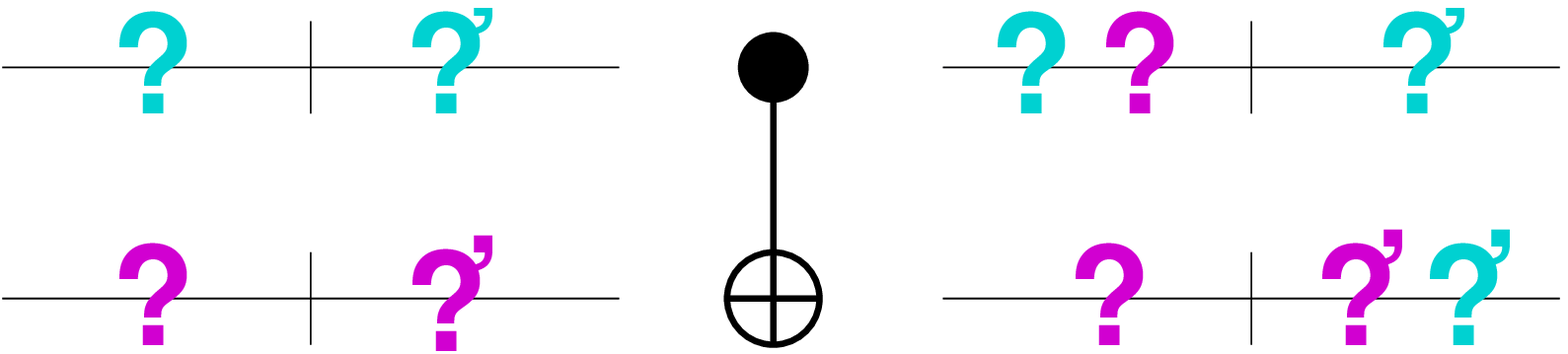}
\end{center}
\vspace{0.2cm}

\noindent We can look in detail at what's going on here by comparing it with the descriptor formalism (\ref{CNOT}). A CNOT operation consists of \emph{copying} descriptor dependencies between qubits, with the $q_z$ component of qubit 1 being copied into the $q_z$ component of qubit 2, and the $q_x$ component of qubit 2 copying back into qubit 1. Another copying operation is the controlled-Z (CZ) gate:

\vspace{0.2cm}
\begin{center}
 \includegraphics[height=2.1cm]{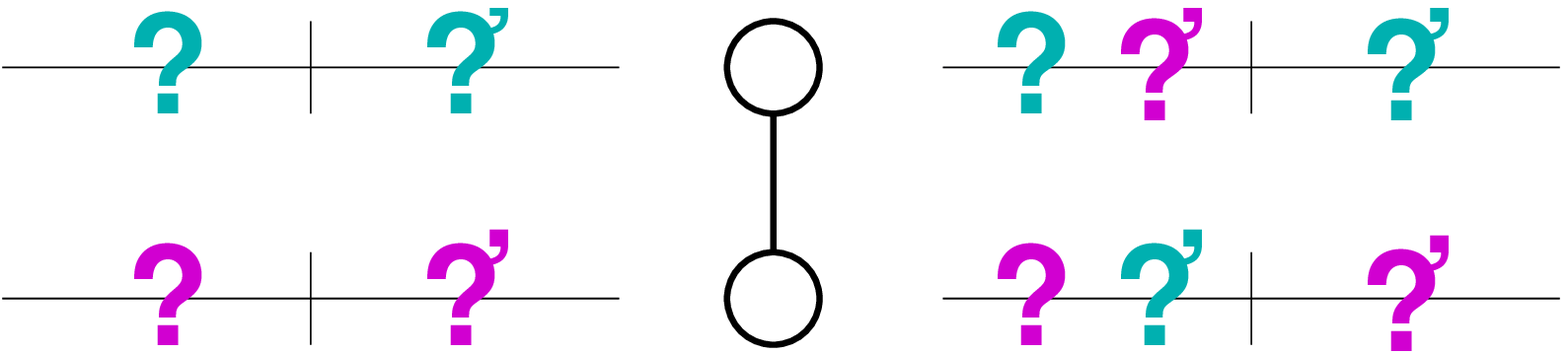}
\end{center}
\vspace{0.2cm}

We now introduce our representation of classical bits using measurement gates. A measurement in the computational basis is represented

\vspace{0.2cm}
\begin{center}
 \includegraphics[height=0.8cm]{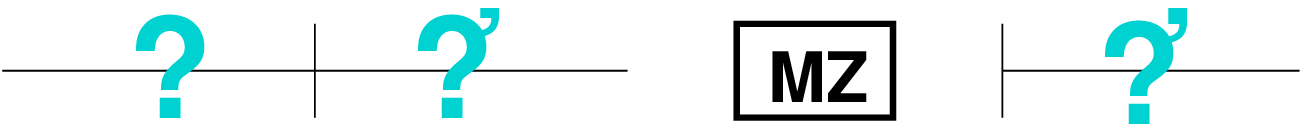}
\end{center}
\vspace{0.2cm}

\noindent Measurements in other bases are implemented by rotations to the computational basis and then a measurement. One specific useful measurement is that in the X-basis, given by a Hadamard and then a measurement. This is represented as a shorthand by

\vspace{0.2cm}
\begin{center}
 \includegraphics[height=0.8cm]{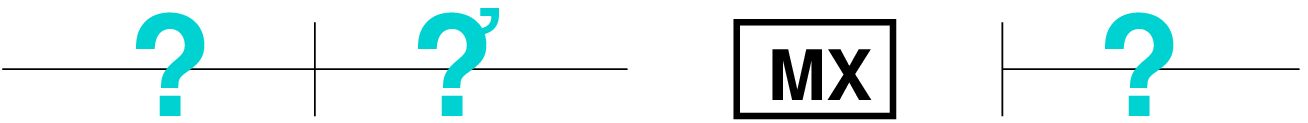}
\end{center}
\vspace{0.2cm}

\noindent In operational terms, the bits given by

\vspace{0.2cm}
\begin{center}
 \includegraphics[height=0.8cm]{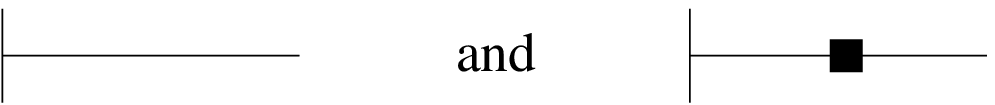}
\end{center}
\vspace{0.2cm}

\noindent are in a definite state in the measurement basis (0 and 1 respectively). Any bit containing a triangle is in a completely random state.

\noindent Bits can now participate in the same single-input NOT gate as is used for qubits:

\vspace{0.2cm}
\begin{center}
 \includegraphics[height=0.8cm]{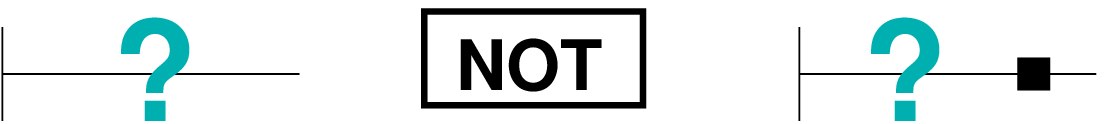}
\end{center}
\vspace{0.2cm}

\noindent and two-qubit gates can take a mixture of bits and qubits as their inputs:

\vspace{0.2cm}
\begin{center}
 \includegraphics[height=10cm]{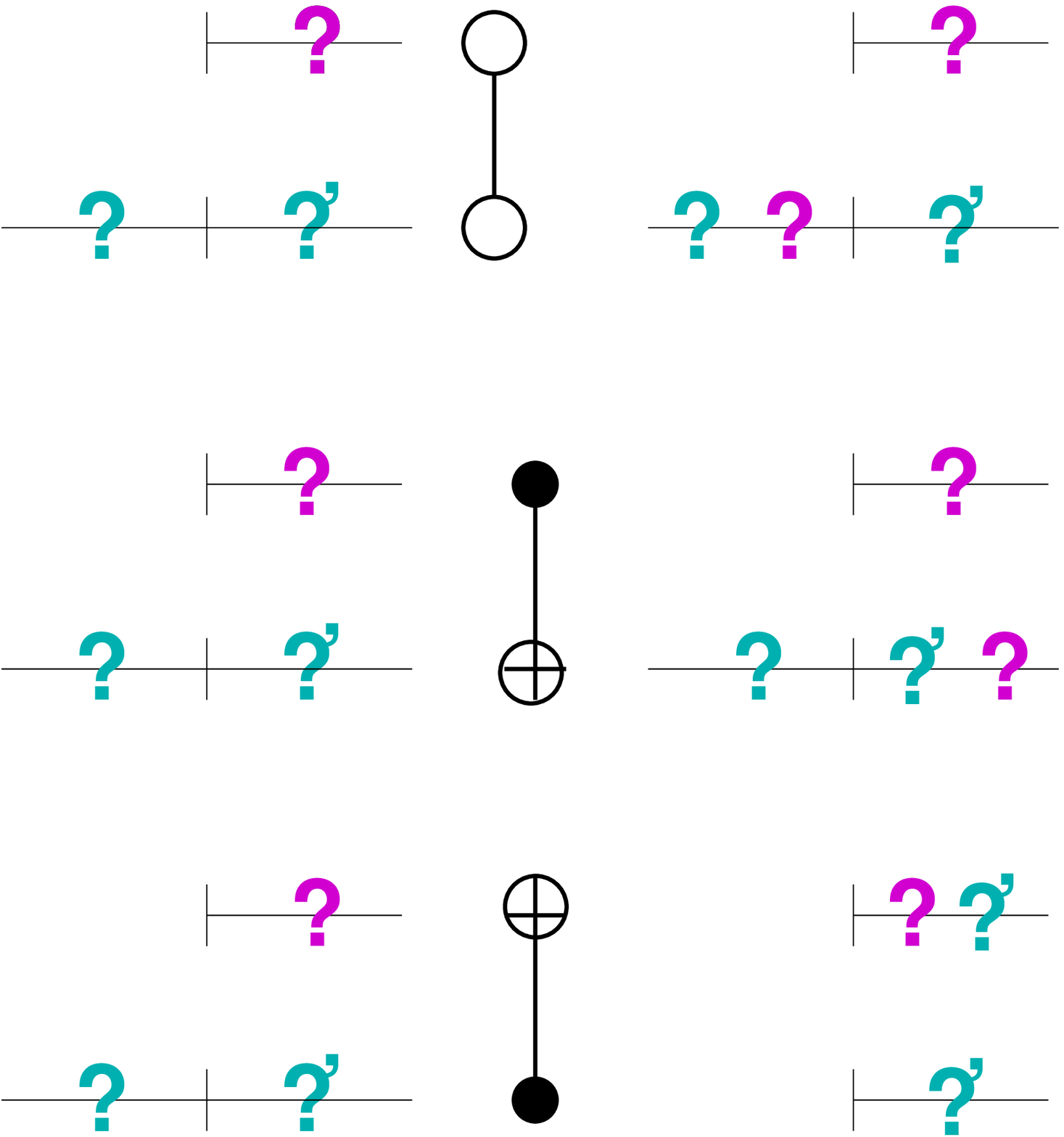}
\end{center}
\vspace{0.2cm}

That is now the calculus defined. What we have given is not fully universal for quantum computing as we have do not have a definition for arbitrary single qubit rotation \cite[p191]{nandc}. We do, however, have a full set of Clifford gates which are widely used in measurement-based computing, and give us an interesting subset of quantum computational behaviour. We can now identify the key components of the symmetric monoidal category structure with strong compact closure used by Abramsky and Coecke \cite{cat}. The \emph{objects} of the system are the contents of the qubits and bits, that is the positions and quantities of the triangles and the rotation boxes (remember that the filled box given by the NOT gate is simply a special type of rotation box). The gates comprise the \emph{morphisms} between different configurations of the qubit/bit contents, and we can easily see how an \emph{identity} could be defined that corresponds to a $\openone$ gate. We have a \emph{symmetry operation} that swaps the state of the qubits (given by the circuit-model SWAP gate):

\vspace{0.2cm}
\begin{center}
 \includegraphics[height=1.6cm]{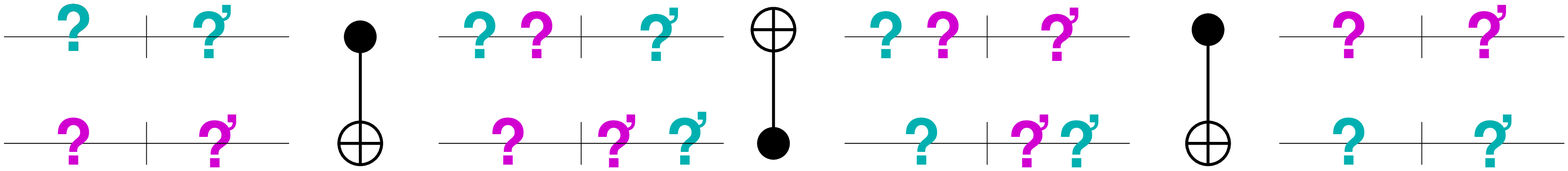}
\end{center}
\vspace{0.2cm}

It is interesting to consider the structure of compound systems in this representation. In contrast to our usual intuitions about the creation of composite systems in quantum mechanics, in this representation a composite system is given by the simple combination of the constituent systems; there's nothing in the joint system that is not present in the individual ones. It is therefore not immediately obvious that we have a tensor rather than cartesian product structure for compound systems. However, when combining systems we not only need to take into account the individual qubit representations, but also the ``triangle'' composition rules, which will be important when gates (including measurement) are performed on the joint system. The asymmetry in the composition rules means that unique subsystem states cannot be recovered from a joint system; the correct \emph{associative connective} for the symmetric monoidal category is indeed the tensor product. Finally, the requirement of strong compact closure is satisfied in systems exhibiting teleportation, we will we demonstrate below. We therefore now have all the elements needed for a graphical representation of quantum mechanics. 


\section{Dependencies and entanglement}

The first important element of quantum computing to consider in this representation is entanglement. As this is the physical outcome of the tensor product structure in quantum mechanics, we can see that it will come from a ``triangle'' composition. To see this in detail, let us look at an entangling Bell operation:

\vspace{0.2cm}
\begin{center}
 \includegraphics[height=2.1cm]{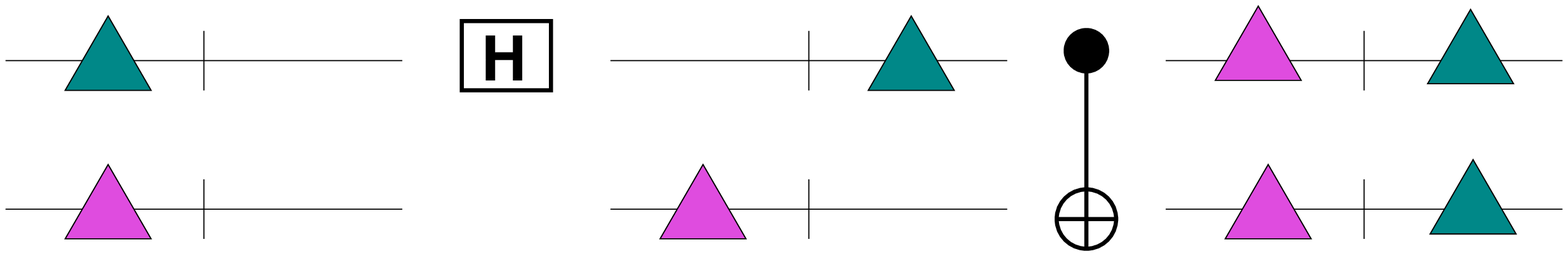}
\end{center}
\vspace{0.2cm}

The first obvious thing to say about this system is that the two qubits finish as identical: each has now incorporated part of the state of the other qubit from the interaction operation. In other words, the state of one qubit now depends on information that it picked up from the other during their mutual interaction. Note that this dependency does not change unless there are local operations performed on that particular qubit; unlike in the standard representation of quantum mechanics, remote operations do not change the state of a qubit. 

So how precisely are these dependencies encoding the entanglement between qubits? As mentioned above, the key to this is the composition rule for triangles. Two qubits are entangled if there is a particular joint measurement which gives a different result from the composition of the individual measurements (eg. \cite[ch2]{nandc}). As the presence of any triangle in a bit means a completely random outcome, entanglement is shown by the ability to use the composition rule to annihilate all triangles from at least one joint measurement outcome bit. Taking the above example of the Bell state, we see that all individual measurements on the two qubits will give random bits. However, we can recover a definite bit in the following way by performing the reverse operation:

\vspace{0.2cm}
\begin{center}
 \includegraphics[height=2.1cm]{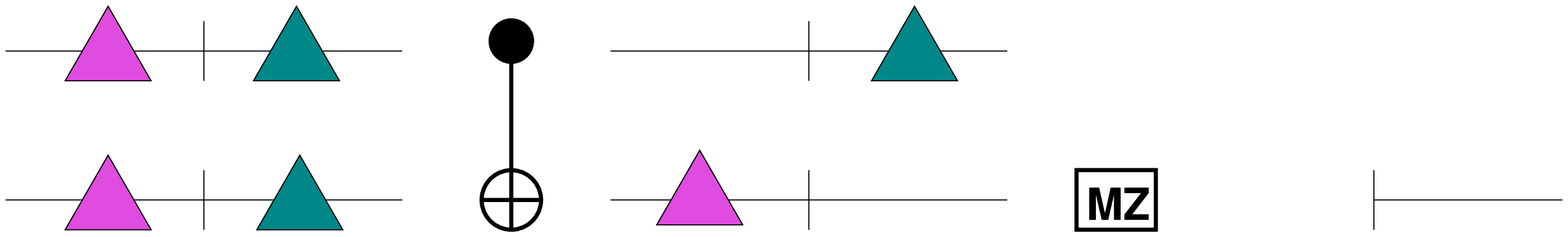}
\end{center}
\vspace{0.2cm}

The true power of the dependencies comes in, however, when they are used to carry information through a computation, and only annihilated at key points. We can see this by turning to arguably the most important measurement-based protocol, teleportation.

In a teleportation protocol, Alice and Bob begin by sharing a pair of qubits in the state $\ket{01} + \ket{10}$:

\vspace{0.2cm}
\begin{center}
 \includegraphics[height=2.1cm]{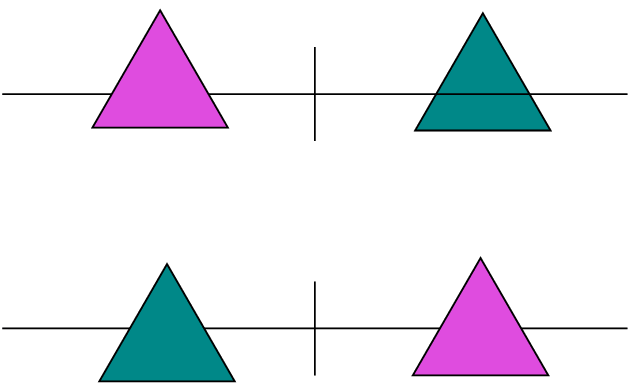}
\end{center}
\vspace{0.2cm}

\noindent (proof that this is indeed the correct representation of the state is left as an exercise to the interested reader).

Alice and Bob are spatially separate, so we can (in this representation) look only at the qubits and bits local to each. Alice now has a qubit in an unknown state that she wishes to teleport. She entangles this with her half of the shared qubit pair, and then performs two measurements:

\vspace{0.2cm}
\begin{center}
 \includegraphics[height=2.1cm]{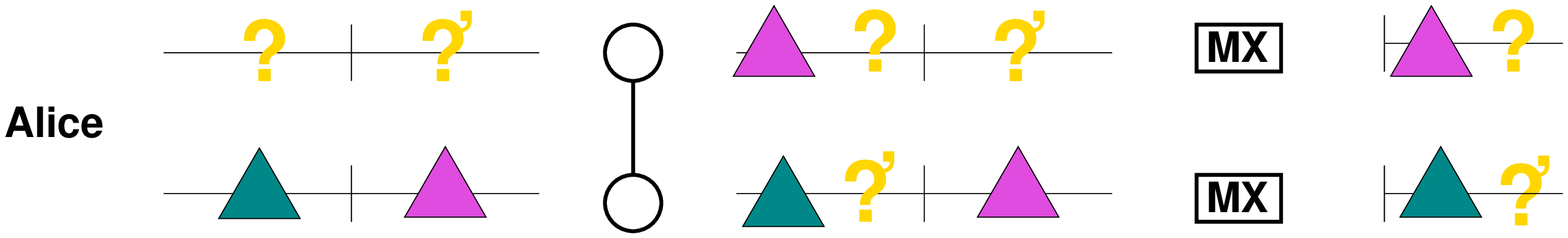}
\end{center}
\vspace{0.2cm}

\noindent These two bits are then communicated over the intervening distance to Bob, using a classical communication channel. Bob performs two operations between the bits and his half of the original qubit pair, and recovers Alice's teleportation qubit:

\vspace{0.2cm}
\begin{center}
 \includegraphics[height=3.6cm]{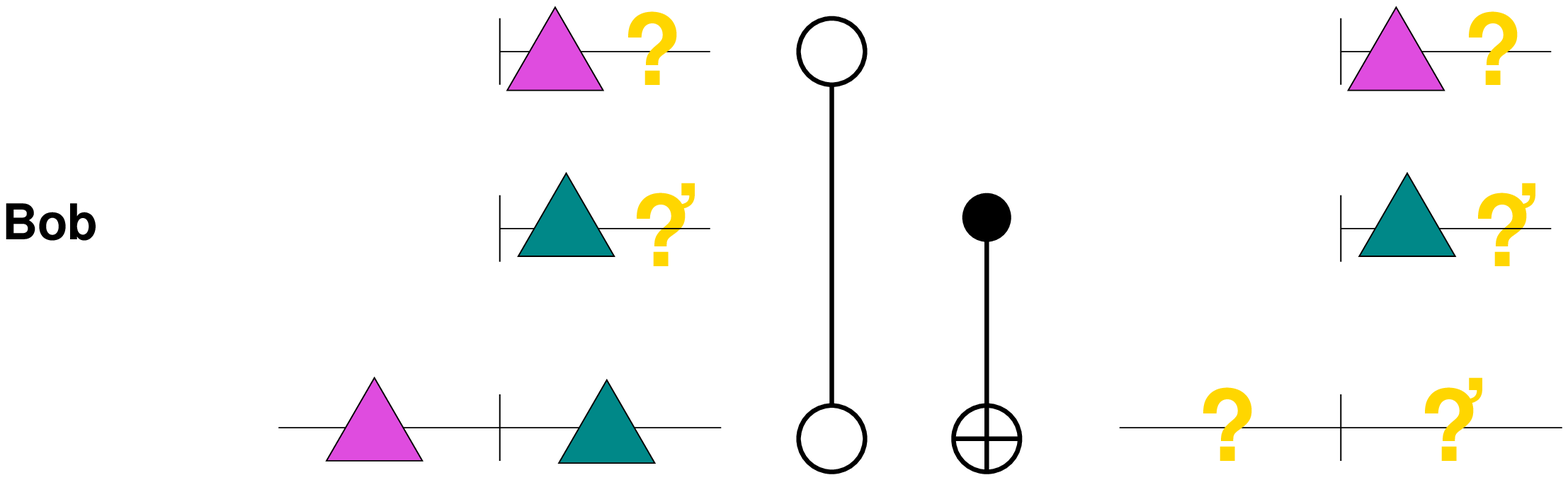}
\end{center}
\vspace{0.2cm}

One immediate consequence of describing teleportation in this representation is that it ceases to become a non-local operation. The state that Alice is sending to Bob does not disappear at one end and reappear at the other; rather, the data contained in the qubit is split into two bits which are then sent from Alice to Bob. An interesting question then becomes what role the pre-shared entanglement is playing in this picture. In the usual model, the role of the entangled qubits is to create a quantum communication channel. Alice's operation that entangles her half of the pair with the teleportation qubit is considered to change the joint state of the shared qubit pair. Bob can then recover the teleported state if he performs certain operations on his half. The classical communication is only needed to tell Bob which operation to do; it does not, in itself, convey any of the quantum information. By contrast, in this picture Alice's operations do nothing to the state of Bob's qubit: it is only when Bob performs the operations between his qubit and the communicated bits that its state changes. Neither quantum nor classical information or dependencies can travel between two points without something physically carrying them.

Rather than acting as a quantum channel, the role of the shared entanglement instead in this picture is analogous to that of an encryption key. The data to be teleported is composed with the ``key'' (the red and green triangles), transmitted, and then decoded at the other end. The sharing of an entangled pair of qubits beforehand then becomes a form of \emph{key distribution}: Bob has an exact copy of the key that Alice is using, generated through the entangling operations that produce the original qubit pair. This is shared ahead of time, and enables him to extract the data contained in the bit pair that is communicated between them. 

This representation of teleportation also significantly changes the emphasis placed on the final operations performed on his qubit by Bob. On a standard understanding, by this point in the process Alice's operations have changed the state of Bob's qubit into a different definite state. This is the original teleportation state, with an additional `Pauli correction' \cite{owqc} (one of four known states). All the classical communication does is tell Bob which correction operation he needs to do in order to recover the teleportation state -- \emph{i.e.} it tells him which of four states he actually has. The correction operation becomes almost an afterthought: Bob has the state, but does not know it yet. The correction operation can therefore be dealt with almost completely informally; the information transmission protocol is complete by that point. In the graphical representation, however, Bob's ``correction operations'' are the only way in which the teleported information gets to his qubit. These ``decoding operations'' are of equivalent importance to Alice's ``encoding operations'', and are entirely integrated into the formalism. This is, in fact, how all such Pauli corrections are represented in the calculus: a Z-correction translates to a CNOT between bit and qubit, and an X-correction is a CZ operation. The feed-forward of classical information therefore becomes an integral part of the \emph{processing} of information in a hybrid computation, rather than what is necessary to read out the results of an already-completed computation.


\section{Classical information}

We have seen that classical feed-forward looks very different in the graphical representation compared with the usual view. Even at a formal level we have defined CNOT and CZ gates which can take both bits and qubits as inputs. This is simply not possible on the standard view, but emerges completely naturally within the graphical calculus. Because this is so different from the usual view of classical information, we are left with the problem of defining what exactly it is that is being transmitted through the classical channel in this representation. The usual view of the classical data is a single measurement outcome, either a 0 or a 1; a detector either clicks or it does not. How then does this relate to the relatively complex objects that are bits in processes such as the teleportation example above?

The first thing to note when trying to understand the role of the classical channel here is that, in a measurement-based computation, the measurement outcomes are never single outcomes in isolation. Future computational actions are changed dependent on these outcomes, so within the logical structure of the computation they carry with them contextual information. A classical measurement outcome does not just mean ``detector X clicks'', for example: it means, in context, ``detector X clicks, this was caused by state Y, and so operation Z should be performed''. This is precisely the interface between the quantum and classical information processing units that is treated informally in standard presentations. As the aim is to fully formalise the entire process, this ``informal'' data will need to be found within the representation itself, rather than added by hand. There are, then, three elements to the representation: quantum systems and processes, classical systems and processes, and the formal treatment of their interaction within a given protocol.

What we find with the graphical representation is that the contextual algorithm information is formalised by being incorporated into the description of the classical channel. This makes intuitive sense as this information concerns the function of a particular bit within the process. In logical terms, then, what is represented by the triangles etc. in a classical bit is not only the 0 or 1 of a straightforward measurement on that bit, but also the information about how that bit will interact with the quantum systems to change the future state of the computation. This in turn will depend on previous states of the computation (for example, in teleportation the states of the bits will depend on the state of the qubit that Alice is trying to teleport). So the description of a bit includes not only the results of measurements on that bit itself, but how those measurements depend on the states of the qubits with which it had previously interacted. 

In general, these dependencies will not have a significant effect on the outcome of measurements on a bit. However, for certain situations such as teleportation, it is possible to arrange the systems so that the bits interact with qubits which contain identical dependencies. In this case, different data can be extracted from the bit than is possible from individual bit measurements. This is done by making sure there are copies of the dependencies used in the systems with which it will interact -- in other words, by pre-sharing entangled qubits.

We can see, then, that the information in the classical channel in this representation fully defines how the bits interact with the qubits. In logical terms, the calculus can be viewed as giving the dependency structure of an algorithm, showing which previous states the future of the computation is dependent on. The representation of classical information therefore also needs to include these dependencies, and to formalise the bit-qubit interaction. 

In physical terms, the dependencies carried by the bits have been generated by entanglement. Shared entanglement gives shared dependencies, and we can also see from the available gates that it is only the entangling operations (CNOT and CZ) that allow us to copy dependencies from one qubit to another. The unusual feature of the classical bits here is that they too can carry the dependencies created by entanglement, and feed this back into the computation by interacting with qubits containing other dependencies.

By carrying these dependencies, the classical channel in this picture has now become part of the computation on a par with the quantum system, rather than merely giving data about the state of the qubits which perform the main computational work. It is the dependencies which enable the computation to happen: the computational resources in this picture are \emph{dependencies} rather than bits or qubits. Bits and qubits become different ways of utilising and transmitting these resources, rather than separate resource entities in their own right. Quantum and classical elements to the computation are unified both by evolving under the same set of actions and interactions (the gates), and also by carrying the same set of fundamental resources. 

Similarities between the processing abilities of quantum and classical elements now come down to how they carry the dependencies of the computation. At the most fundamental level in the calculus, qubits have two separate components whereas bits have one. When these are acted on by identical gates, there are therefore differences in how bits and qubits appear to act, even though the dependencies are interacting in exactly the same way in all cases. One consequence of this fundamental similarity is that measurement-based computation is possible at all: classical elements can be used in the computation because they carry the same resources as quantum elements. If at some point during a quantum computation these only need to be transmitted individually rather than in pairs, then classical bits rather than qubits can be used. Moving between quantum and classical parts of a computation now becomes a question of how we wish to implement the underlying dependency structure of an algorithm. If we choose to use only qubits then we have a coherent computation with classical read-out at the end, but if we choose to send some dependencies through the system individually then we have a measurement-based implementation. Fundamentally, these are the same type of information processing, and the graphical calculus gives us a visual and intuitive method of translating between the two types of computation.

This basic similarity between quantum and classical information can also show the apparently fundamental differences between how information is stored and processed in bits and qubits. Arguably the most important difference between them is the persistence of information: classical bits can be cloned and deleted, whereas qubits are subject to the tight constraints of the no-cloning and no-deleting theorems. From the structure of the graphical calculus we can note the equivalent rules for the dependency information, which turn out to differ from both: \emph{dependencies can be cloned and annihilated but not deleted}.

The cloning of dependencies is straightforward; this is what an entangling operation does between two (qu)bits. Entanglement simply is the sharing of identical copies of dependency information in this picture. The interesting question then becomes why we can't clone two dependencies from the left and right hand sides of a qubit at the same time, but we can just for one -- in other words, why a qubit cannot be cloned but a bit can. The key to this lies in the entangling operations in this representation, in particular the CNOT operation. The CNOT operation here is symmetric between the control and target, up to a left/right symmetry on the inputs. The dependencies from the right-hand side of the control are copied into the target, and the target reacts back by copying its left-hand dependencies back to the control. Only one side of a qubit can be copied to another during an entangling operation -- in order for both sides to be copied, two operations would be needed. The CNOT operation can be thought of as a ``perfect measurement'', and so this can be described as a measurement not being possible without a back-reaction from the measuring system onto that being measured. However, after one interaction the state of the system is no longer the same as before the measurement, so trying to extract the second set of dependencies will give the original set but composed with the dependencies picked up during the measurement. So if the information containing system has only one set of dependencies then it can be cloned, but if it contains two then the necessary back-reaction from any entangling operation means that it cannot be duplicated. Bits can be cloned, qubits cannot, but both are a consequence of how dependencies are copied between systems.



\section{A local information flow}

We now turn to the second motivation for developing the graphical calculus, that of giving a logically and physically local notion of information flow through a hybrid computation, with the flow respecting standard notions of causality (such as action only into the future). There are a number of conditions that such a flow would have to satisfy in order to be considered as local, and we will look at these in turn.

Firstly, there is the question of what exactly is representing the `information' that is `flowing' in this picture. What is it that gets tracked through a computation? As we would expect, what are tracked are the resources for the computation -- in this representation, information flow is the flow of dependency information through both bits and qubits.

The simplest locality condition on this flow is that the representation of an individual system must give all the information available for the evolution of that system. This is indeed what the individual calculus elements do, by construction. The next condition is much stronger, and is not satisfied by the usual state descriptions in quantum mechanics. This is the condition that there is no global state information that is not present in the local state. In standard quantum mechanics, the product of individual state descriptions of some set of systems (the reduced density matrices) is in general different from the description of the global state (the full density matrix). By contrast, in the graphical representation given here, there is no information contained in a joint state than is not already present in the individual states. There is no separate representation of joint system states in the calculus -- the nearest to such an idea would be how two or more systems evolve under the action of a joint gate. Individual systems retain their individual representations even under entangling operations, and giving the set of individual states in a system gives the complete evolution of that system as a whole.

As well as this ``mathematical locality'', we also want our information flow to satisfy ``physical locality'' conditions. The most important of these is that local operations change only local states. This again is something that is usually violated in standard representations of quantum mechanics, but arises naturally in the calculus. This is one of the differences that we saw during teleportation. In the standard picture, Alice's local operations on her qubits change the state of Bob's qubit non-locally. By contrast, in the graphical representation, Alice's operations only act locally, as do Bob's. Information flows between them down the classical channel, rather than jumping from one to the other.

The final condition is that not only do local operations only change local states, but that the \emph{only} way to change the state of a system is by locally operating on it. This is true in the graphical calculus by construction: the only morphism on an object is a gate, and the only gates that have been defined are those that act locally.

We therefore have a graphical representation of a flow of information through a computation that is represented using objects local to the systems being described, and which change only when acted on locally. This flow is not only local but also \emph{causal}. Information contained within a system is only changed by actions on that system, and these actions only occur in one temporal direction. Information flows forward in time through a computation and never backwards, either logically or physically. 



\section{Conclusion}

We have developed here a graphical calculus for information flow in hybrid information processing systems. This is a framework for information processing that integrates both quantum and classical elements into the same formalism, and explicitly defines the interface between bits and qubits using the same interactions as between qubits alone. This allows a fully formalised description of measurement-based computations, including a full description of the role played by the classical information channel. This leads to new descriptions of hybrid computational situations such as teleportation, with information previously considered to be entirely quantum being transmitted through the classical channel.

Quantum and classical resources in a computation are unified in this representation as instances of the underlying dependency resources. These dependencies are the basis for computation in this model, and their transmission and processing during a protocol demonstrate the flow of information. This flow model is local and causal, and defines both the logical structure of an algorithm and the physical flow of information within it. One interesting question is the connection between the information flow given by this representation and the flow notions defined in \cite{cat,elhamflow}, and whether this physical flow underpins these logical flow ideas.

Finally, we note that we now have the basis for a formal model of computation for both quantum and classical elements that can enable us to make direct comparisons between the two. Such comparisons could include resources use, including how quantum and classical resources can be interchanged for different implementations of a particular algorithm. Such a formal computability model would have its basic elements as the dependencies which are contained in both quantum and classical systems, and would allow the calculation of, for example, the number of classical bits and gates needed to implement a known quantum algorithm. This in turn is an important step towards giving a formal definition of the relative processing powers of quantum, classical and hybrid computing.

\section*{Acknowledgements}

We would like to thank Keith Harrison for prompting the construction of the calculus, and for critical feedback during its development. We also acknowledge fruitful discussions with Damian Markham, Kae Nemoto and Viv Kendon. This work was supported by European projects QAP and HIP.



\bibliographystyle{plain}
\bibliography{biblogMAIN.bib}

\end{document}